\documentclass[12pt,preprint]{aastex}

\begin{document}
\newcommand{\msun}{M_{\odot}}
\newcommand{\kms}{\, {\rm km\, s}^{-1}}
\newcommand{\cm}{\, {\rm cm}}
\newcommand{\gm}{\, {\rm g}}
\newcommand{\erg}{\, {\rm erg}}
\newcommand{\kpc}{\, {\rm kpc}}
\newcommand{\mpc}{\, {\rm Mpc}}
\newcommand{\seg}{\, {\rm s}}
\newcommand{\kev}{\, {\rm keV}}
\newcommand{\hz}{\, {\rm Hz}}
\newcommand{\sr}{\, {\rm sr}}
\newcommand{\etal}{et al.\ }
\newcommand{\yr}{\, {\rm yr}}
\newcommand{\eq}{eq.\ }
\newcommand{\lya}{Ly$\alpha$\ }
\newcommand{\hi}{\mbox{H\,{\scriptsize I}\ }}
\newcommand{\hii}{\mbox{H\,{\scriptsize II}\ }}
\newcommand{\hei}{\mbox{He\,{\scriptsize I}\ }}
\newcommand{\heii}{\mbox{He\,{\scriptsize II}\ }}
\newcommand{\cii}{\mbox{C\,{\scriptsize II}\ }}
\newcommand{\ciis}{\mbox{C\,{\scriptsize II}${}^{\ast}$ }}
\newcommand{\nhi}{N_{HI}}
\def\arcsec{''\hskip-3pt .}

\title{
The Stryngbohtyk Model of the Universe:
A Solution to the Problem of the Cosmological Constant}
\author{Jordi Miralda-Escud\'e}
\affil{Institut de Ci\`encies de l'Espai (IEEC-CSIC)/ICREA, Barcelona}
\email{miralda@ieec.uab.es}

\begin{abstract}
  Astronomical observations have shown that the expansion of the
universe is at present accelerating, in a way consistent with the
presence of a positive cosmological constant. This is a major puzzle,
because we do not understand: why the cosmological
constant is so small; why, being so small, it is not exactly zero; and
why it has precisely the value it must have to make the expansion start
accelerating just at the epoch when we are observing the universe. We
present a new model of cosmology, which we call the stryngbohtyk model,
that solves all these problems and predicts exactly the value that the
cosmological constant must have. The predicted value agrees with the
observed one within the measurement error. We show that in the
stryngbohtyk model, the fact the cosmological constant starts being
important at the present epoch is not a coincidence at all, but a
necessity implied by our origin in a planet orbiting a star that formed
when the age of the universe was of the same order as the lifetime of
the star.
\end{abstract}

\keywords{cosmology: theory -- shape of the universe -- anthropic principle
is rubbish -- strings, branes and funnels}

\section{Introduction}

  We have learned over the last decade that the expansion history of our
universe is described by the Friedmann equation derived from the General
Theory of Relativity with the addition of a cosmological constant, which
has just the value that makes the expansion start accelerating around
the present epoch. Evidence for this strange result has come from the
observations of the detailed shape of the power spectrum of fluctuations
in the Cosmic Microwave Background (Spergel \etal 2006 and references
therein), Type Ia supernovae (Riess \etal 2006 and references therein),
and other confirming evidence such as the values of the Hubble constant
and the age of the oldest known stars, and the evolution of galaxy
clusters. If the accelerated expansion is interpreted as the result of a
component with negative pressure $p=w \rho c^2$, present observations
show that $w$ is consistent with a constant value of $-1$, corresponding
to the simple case of a cosmological constant.

  The cosmological constant has an interesting history. Einstein first
added it to the equations of General Relativity in order to obtain a
static, closed universe, where matter's attraction and the cosmological
constant repulsion exactly balance each other. After the expansion of
the universe was discovered by Hubble in 1929, the cosmological constant
was discarded and astronomers took to the task of measuring the only two
parameters that were thought to be left to measure, the Hubble constant
and the matter density (although the Steady State model of the universe
was proposed as an alternative until it was observationally ruled out).
After seven decades of controversy, astronomers finally managed to
measure the evolution of the expansion rate of the universe around the
turn of the century, and they had to agree that there is, after all,
evidence for the kind of accelerated expansion that is produced by the
cosmological constant, in a flat universe. Now, just like it happened
after Hubble's discovery of the expansion, all the astronomers are
getting excited about measuring more details of this acceleration and
learning the fate of the universe.

  It is curious that, despite the absence of any really new theoretical
developments to understand the reason for the accelerated expansion, and
despite the perfect agreement of the observations with the most simple
possibility of a cosmological constant, countless papers are being
published on the possibilities to produce accelerated expansion: all
types of modifications of gravity, as well as hypothetical
components with negative pressure that have been named ``dark energy''.
The exhilaration has reached such an extreme that cosmologists are heard
these days talking about ``dark energy'' as if this were a real, already
detected substance.

  As a note, the name dark energy for a component driving the
acceleration is particularly bad among all the bad terminology that
astronomers have made up, because Einstein discovered that
\begin{equation}
 E=mc^2 ~,
\end{equation}
(this equation is written here in case anybody had forgotten it),
so everything in the universe is energy (and the name ``dark stuff''
would be no worse than dark energy). Moreover, dark means something
that absorbs the light, whereas something that lets all the light go
through without interaction should be called transparent, or invisible
(which means detectable only through gravity, because Einstein found
that nothing can be invisible to gravity). Perhaps a better name would
be invisible tension: the distinguishing property of a component of the
universe accounting for the acceleration would have to be, after all,
its negative pressure.

  The detection of the present acceleration of the expansion poses
a very deep puzzle for cosmology and for all physics. What this
observation is telling us is that the cosmological constant has a
value that is 123 orders of magnitude smaller than its only natural
magnitude one can think of, the Planck density. So, the terrible
questions we face are: Why is the cosmological constant so small
compared to this natural value? (this is the question we had before,
on which we have made no progress); why, being so small, is the value
not exactly zero? (so, it is not enough to have some symmetry that makes
the cosmological constant be zero, but some small correction is needed);
and why, being not exactly zero, it just
happens to have exactly the value that makes its density similar to
the average matter density, at the time when a biological species that
may be more or less intelligent and has appeared in some random planet
starts to wonder about the universe? (Well, at least some of the
individuals in the species do; most care only about money, sex,
football, and management of power).
In fact, the epoch when the cosmological constant is
exactly half of the total energy density is at redshift $\sim 0.3$,
very close to the present. The problem is so hard to deal with that
some cosmologists, losing all shame, have even appealed to
anthropic principles.

  This paper presents a new model of cosmology, the stryngbohtyk
\footnote{The etimology of the word stryngbohtyk comes from the Catalan
language, from the word ``estramb\`otic'', which means something that
is out of the ordinary in an extravagant and laughable way.} model
of the universe. It solves all the problems associated with this
detection of the accelerating expansion: it predicts exactly the value
the cosmological constant must have, and, you will be amazed to find
out, the value agrees with the measured one within the error. Moreover,
it will be shown that the similarity of the predicted value with the
present matter density is not a coincidence at all, but is a necessity:
whenever an intelligent species arises in a planet at a time when the
age of the universe is of the same order as the lifetime of its host
star, the epoch when the acceleration starts must be roughly
of the same order as the epoch at which the universe is observed.

\section{The origin and shape of the universe in the stryngbohtyk model}

  All the present cosmological data is explained by the structure
formation model of Cold Dark Matter, which postulates that the dark
matter is made of collisionless particles or objects that have a
negligible
initial velocity dispersion, and that there are Gaussian, adiabatic
primordial perturbations with a nearly scale-invariant power-law power
spectrum. This was initially postulated for reasons of simplicity. The
amazing thing is how well this simple model fits the very detailed
measurements of the CMB by the WMAP mission, as well as various other
astronomical observations of large-scale structure, once the
cosmological constant is included.

  In this context, present cosmology has come to be dominated by the
concept of inflation, which essentially proposes the very
naive and generic idea that the primordial perturbations were causally
generated in the early universe and then inflated out of the horizon by
an accelerated expansion similar to the one that is starting at present
(although with a much higher Hubble rate). Then, this idea is used to
attribute all the success of the Cold Dark Matter model matching the
observations to the inflationary ideology, hence greatly inflating
inflation's merits. Moreover, inflation has the advantage of making a
lot of predictions, which can be changed whenever they are not matched
by observation. In this way, the concept of inflation becomes an
eternally self-reproducing one in the minds of cosmologists.

  Despite inflation's great success, it is always worth considering
alternatives, like the cyclic model (see Steinhardt \& Turok 2005 and
references therein). The new model we present in this paper, called the
stryngbohtyk model, takes a further step in sophistication. Like in the
cyclic model, the universe in the stryngbohtyk model is a brane that is
contained in some higher dimensional space that is called the bulk, and
the particles that we see are strings that are confined to the brane, and
can only interact with other strings in the brane.
However, whereas the cyclic model has two flat branes separated by a
short distance across the bulk, which hit each other at the end of each
cycle starting a new Big Bang (after a period of accelerated expansion
of the branes in which the entropy of the old cycle gets diluted), the
stryngbohtyk model has only one brane which is closed. After the brane
undergoes a period of exponential expansion, then instead of having two
plane-parallel, infinite branes which hit each other nearly
simultaneously everywhere and bounce back, the closed brane of the
stryngbohtyk model hits itself at some singular point, or string. When
the brane hits itself, it can rupture and get reconnected, and develop a
topological hole. After the collision, the universe bounces back and
starts exponential expansion again, generating primordial perturbations
until reheating occurs, making everything just like inflation (whereas
in the original cyclic model the Big Bang phase starts after the brane
collision, and the perturbations are created before the collision).
This is good news, for as inflationary cosmologists say, any model
explaining the flatness and horizon and the rest of you-know-which
problems is like inflation, or else it must be wrong.

  For example, a brane that is initially like a two-dimensional
sphere may contract, becoming some sort of pancake and finally
hitting itself at one point. At the collision, the brane
gets ripped up and reconnected with a topological hole, undergoing a
transition that converts it into a doughnut. Then, any strings that 
happened to be lying around the point of rupture at the instant of the
collision are trapped and forced to expand as the hole of the doughnut
grows, after the collision of the brane. This can happen similarly
with more dimensions, for example in a three-dimensional brane hitting
itself along a string.

  But because of the special symmetries of string theory, there need to
be a total of nine spatial dimensions, three of which are in the brane
and are able to get stretched, and the others may be the bulk or may be
dimensions that remain wrapped up at the Planck scale; and at the same
time, in order for the particle properties and gauge interactions to
come out right, the universe must have two holes, which means the brane
has collided with itself in two places, in which case one of the many
possible vacua of string theory is the one that is right for us. All the
details cannot be explained here; but in any case, because the
stryngbohtyk model is based on string theory, it is a theory of
everything, that is to say, it can explain everything that has ever
been, is, and will ever be.

  So, the universe is a brane that is like the surface of a doughnut
with two holes, not just one. And this can be thought of as the shape of
a funnel, where the brane is the surface and there is a closed bulk
(the plastic or aluminum that makes a funnel) and an open bulk (the
space around the funnel). The closed bulk may be very thin so that
locally it looks like the two branes separated by a small distance, like
in the cyclic model. Some
strings contained in the brane may have been trapped around either one
of the holes when the collisions occurred, and the universe is full of
them with all possible combinations. Figure 1 illustrates the shape of
the universe in the stryngbohtyk model; our brane is both the inner and
outer surface of the funnel (these surfaces should join smoothly at the
top and bottom of the funnel even though it is not shown in the figure).
One hole is the bulk inside the funnel and the other is at the handle.
So the trapped strings go either around the funnel, or around the
handle.

\begin{figure*}[t]
\plotone{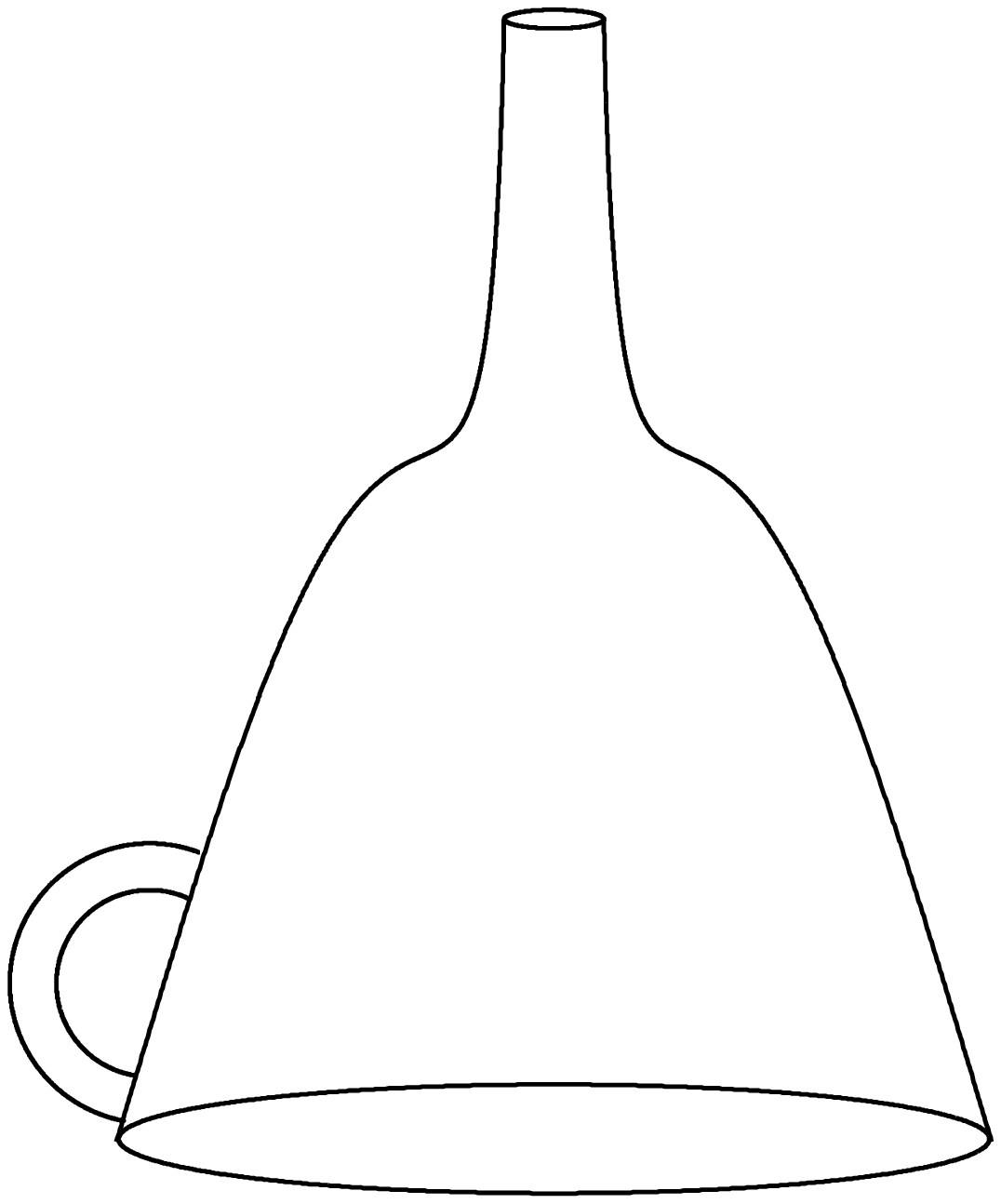}
   \caption{Shape of the universe according to the stryngbohtyk model.
Our universe is a brane with the shape of the surface of the funnel
(both inner and outer surface, connected on the high curvature regions
at the edges, not shown in the figure), and the particles and forces
we observe are strings confined in the brane. There is an inner closed
bulk and an outer bulk, and the universe has two topological holes. For
better inspiration to think on the stryngbohtyk model, it is recommended
to place an object as shown in the figure on top of one's head.}
	 \end{figure*}

\section{Prediction for the cosmological constant}

  It turns out that, due to a special symmetry that arises in the
stryngbohtyk model, there is a cancellation of all the contributions to
the vacuum energy density coming from the usual strings in the brane
with the strings that are trapped around a hole.
If it were not for the topological holes, the cosmological constant
would not get cancelled and it would be of order the Planck density.
But because of the holes, the cancellation occurs and the universe can
exist for much longer than a Planck time.
After very long calculations, one finds that even with the holes there
are high-order terms for the vacuum energy density which do not cancel, due to 
particles of spin $1/2$ in the three families, which come out
depending on the rest-mass of each particle in the three
families, all multiplied together, like this:
$(m_{i1} m_{i2} m_{i3})^n/n!$, where the number says which
family the particle belongs to, $n$ is the number of topological holes
in the universe, and the $i$ represents the type of
particle, and we use Planck units. So, the most important contribution
to the cosmological constant comes from leptons. The neutrinos have
much smaller masses and their contribution is negligible, and the
three colors of the quarks make them contribute a term going as the
cube power of their multiplied masses, so they are negligible too.

  So, the predicted value of the cosmological constant is
$(m_e m_{\mu} m_{\tau})^2/2$, where the masses are those
of the electron, muon and tau particle, and we have used $n=2$. 
With all the units put back into the equation, the stryngbohtyk
prediction for the cosmological constant is
\begin{equation}
{\rho_{\Lambda} \over \rho_{Pl} } = { 3H_0^2 \Omega_\Lambda \over
8 \pi G \, \rho_{Pl} } =
 { (m_e m_{\mu} m_{\tau})^2 \over 2 m_{Pl}^6 } ~ .
\label{strp}
\end{equation}
Here, $H_0$ and $\Omega_{\Lambda}$ are the things familiar to
astronomers, the Hubble constant and the density of the cosmological
constant in units of the critical density, and $m_{Pl}$ and $\rho_{Pl}$
are the Planck mass and Planck density:
$m_{Pl} = (\hbar c/G)^{1/2} = 2.177 \times 10^{-5} \, {\rm g}$,
and $\rho_{Pl} = c^5 /(\hbar  G^2) = 5.16 \times 10^{93}
\, {\rm g}\, {\rm cm}^{-3}$. The particle masses are (e.g., Eidelman
\etal 2004), $m_e = 9.109\times 10^{-28}\, {\rm g}$, 
$m_{\mu} = 1.883\times 10^{-25}\, {\rm g}$, 
$m_{\tau} = 3.168\times 10^{-24}\, {\rm g}$, so the predicted value
of the density of the cosmological constant in Planck units is
\begin{equation}
  { (m_e m_{\mu} m_{\tau})^2 \over 2 m_{Pl}^6 } =
  1.388\times 10^{-123} ~ .
\end{equation}
The largest error of the particle masses is for the $\tau$ particle,
which implies an error on this prediction of less than one part in a
thousand. The measured cosmological constant, using values
$H_0 = 100 h_0 \kms/\mpc^{-1} = 73 \pm 3 \kms/\mpc^{-1}$,
$(1-\Omega_{\Lambda}) h_0^2 = 0.13\pm 0.01$
(Spergel \etal 2006), is
\begin{equation}
{ 3H_0^2 \Omega_\Lambda \over 8 \pi G} \, \rho_{Pl}^{-1} =
  (1.48\pm 0.16) \times 10^{-123} ~ .
\end{equation}
The good news for astronomy is that there is now an added value to
measuring the Hubble constant more accurately, namely to see if the
stryngbohtyk prediction holds up.


\section{Solution to the coincidence problem}

  So, the stryngbohtyk prediction turns out to work, at least for now.
But, how is the problem of the
coincidence explained? Do we simply have to assume that there is a
theoretical prediction for the epoch when the acceleration starts, and
we happen by chance to live at this epoch, or can we understand this
thing better?

  Decades ago, Professor P.A.M. Dirac also noticed another funny
coincidence in our universe (Dirac 1937). The ratio of the electric to
the gravitational force between an electron and a proton is
\begin{equation}
 R_{ge} = {e^2 \over Gm_p m_e } = 2.27\times 10^{39} ~.
\end{equation}
Also, the number of baryons within the observable horizon is equal to
(using for now an Einstein - de Sitter universe for simplicity, with
Hubble constant $H_0= 73 \kms\mpc^{-1}$ and $\Omega_b = 0.04$)
\begin{equation}
 N_p = {4 c^3 \Omega_b \over H_0 G m_p } = 1.63 \times 10^{79} ~.
\label{barn}
\end{equation}
Professor PAM noticed with curiosity that these big numbers of the
universe, one related to fundamental physics and the other to the
epoch when we are observing the universe, seem to be roughly related
as $N_p \sim R_{ge}^2$, and this was a strange coincidence indeed.

  A few decades later, Professor Bob Dicke pointed out that this was
actually no coincidence (Dicke 1961). Given the facts that the
(supposedly) intelligent beings observing the universe arose on a planet
supplied with the light from a star, and that the time it took for these
beings to evolve is not much smaller than the stellar lifetime, it is
not surprising that the first opportunities for these beings to appear
in the universe would occur when the age of the universe is of the same
order as the stellar lifetime. This is in any case a coincidence that we
know is true for us: the Sun's lifetime is $10^{10}$ years, roughly
the same as the present age of the universe. So this must imply a
relation between some fundamental constants and the present age of the
universe.

  To derive this relation, we note first that in a star in hydrostatic
equilibrium that has a characteristic internal pressure $p$, with
contributions from gas pressure $p_g = (1-\beta) p$ and from
radiation pressure $p_{rad} = \beta p$, a fraction $\beta$ of the
hydrostatic support against gravity needs to be provided by the
radiation pressure. Therefore, if the opacity is dominated by electron
scattering, the luminosity of the star needs to be
$L \sim \beta L_{Edd}$ (where $L_{Edd}$ is the Eddington luminosity),
because by definition when the luminosity is equal to the Eddington
one, the radiation pressure exactly balances gravity. In general, a star
may have other contributions to the opacity (e.g., free-free and
bound-free transitions with heavy ions), and then the luminosity will be
further reduced. So in general, the luminosity of a star is
$L = \ell L_{Edd}$, where $\ell = (\kappa_e/\bar\kappa) \beta $,
$\kappa_e$ is the electron scattering opacity, and $\bar \kappa$ is a
sort of average effective opacity in the stellar interior. In general,
$\ell$ increases with stellar mass. For very massive stars $\ell$ is
close to one, and for low-mass stars $\ell$ is
small (for the Sun, $\ell \simeq 10^{-4.6}$, and for an object at the
borderline between stars and brown dwarfs, $\ell \sim 10^{-7}$),
so its value depends on complex biology determining the mass of the star
that is most appropriate for harboring a planet with life. 
The Eddington luminosity is given by
\begin{equation}
 L_{Edd} = {4\pi c G \mu_e \over \sigma_e } \, M =
 {3c^3 G \mu_e m_e^2 \over 2 \hbar^2 \alpha^2} \, M ~,
\end{equation}
where $\mu_e$ is the mean mass per electron (equal to $1.2 m_p$ for
the fully ionized primordial mixture of hydrogen and helium), and
$\alpha$ is the fine structure constant. If the star converts a
fraction $\epsilon$ of its rest-mass energy into radiation over its
lifetime, then its lifetime is
\begin{equation}
 t_s = {M \epsilon c^2 \over L} = {2\alpha^2 \epsilon \over 3 \ell }\,
 {m_{Pl}^3 \over \mu_e m_e^2} \, t_{Pl} ~,
\label{starl}
\end{equation}
where $m_{Pl}$ and $t_{Pl}$ are the Planck mass and Planck time.

  Now, the number of baryons in the universe (eq.\ \ref{barn}) can be
reexpressed as
\begin{equation}
 N_p = {\Omega_b \over H_0 t_s}\, {4c^3 t_s \over G m_p } = 
 {\Omega_b \over H_0 t_s}\, {8\alpha^2 \epsilon \over 3 \ell } \,
 { m_{Pl}^4 \over \mu_e m_e^2 m_p } =
 {\Omega_b \over H_0 t_s}\, {8 \epsilon \over 3 \ell } \,
 { m_p \over \mu_e } \, R_{ge}^2  ~.
\end{equation}
The dimensionless numbers relating $N_p$ and $R_{ge}^2$ can naturally
be expected to be not far from unity. Hence this shows that the
coincidence of the big numbers of Professor PAM is actually not
surprising, but it is simply a consequence of living next to a star that
has lived and will live for a time not so different from the present
age of the universe.

  But now, we see that the reason the acceleration of the expansion
is starting just today is the same one. The time when the universe
expansion starts accelerating, using equation (\ref{strp}), is
\begin{equation}
 t \simeq \left( {3\over 8\pi G \rho_{\Lambda} } \right)^{1/2} =
 \sqrt{3\over 4\pi} \, {m_{Pl}^3 \over m_e m_{\mu} m_{\tau} } \, t_{Pl} ~. 
\end{equation}
The ratio of this time to the lifetime of the star (eq.\ \ref{starl}) is
\begin{equation}
 { t \over t_s } \simeq {3 \sqrt{3}\, \ell \mu_e m_e \over 
 4\sqrt{\pi} \alpha^2 \epsilon m_{\mu} m_{\tau} } ~.
\label{trat}
\end{equation}
This solves the coincidence problem of the cosmological constant.
That is to say, the strynbohtyk prediction that the cosmological
constant density scales as the sixth power of the ratio of typical
particle masses to the Planck mass (eq.\ [\ref{strp}])
implies that the coincidence of the age of the universe
at the time the acceleration starts with the present age is a
necessity. Of course, to make this coincidence more outstanding, it
is still necessary that the combination of dimensionless constants
appearing in equation (\ref{trat}) turns out to be close to unity. The
value of these constants depends on highly complex and diverse physics:
the strength of the electromagnetic interaction, the ratio of the mass
of leptons to the proton mass, stellar physics, and the complex biology
that affects which stellar mass is most appropriate for life. However,
because these constants are not different from unity by too many orders
of magnitude, it makes us feel better to say that the combination in
equation (\ref{trat}) just happens to be close to unity by pure
coincidence, than in the case when we did not have the big numbers of
the universe cancelling out.

\section{Discussion}

  Such is the bewilderment caused by the detection of an
accelerated expansion, that most astronomers in the world, leaving
aside any other more mundane astrophysics,
are focusing their efforts and proposals into methods
to find out something else about ``dark energy'', whatever it may
be.

  The stryngbohtyk model has been proposed in this paper, in which
the universe is a brane with funnel shape in the nine-dimensional
space of string theory, where some dimensions got curled up at the
Planck scale to make all the observed particles and gauge interactions
from strings that are confined to the funnel brane (with the familiar
three extended dimensions), which lives in the bulk (which has the rest
of the dimensions), after selection of one among many possible vacua.
As in the cyclic model, there is a brane collision giving rise to the
Big Bang. In the cyclic model, there are two
plane-parallel, infinite branes that collide. But in the stryngbohtyk
model, the brane collides with itself, the collision can puncture
the brane changing its topology, and strings get trapped around the
created hole and are forced to expand as an epoch of inflation
gets going. Today, we live in a very small patch of the brane with
funnel shape and we cannot realise the true topology of the universe.
The primordial perturbations are homogeneous as in inflation, and
there are no monopoles or primordial black holes. This is, by the way,
truly a pity, because if evaporating black holes could be discovered,
Stephen Hawking would get his Nobel prize.

  Curiously, the stryngbohtyk model makes a prediction,
obtained with stryngbohtyk reasoning, of the exact value of the
cosmological constant, in terms of the lepton masses, which matches
the observation.

  Not only that. In fact, any model, stryngbohtyk or not, in which the
cosmological constant has a reason to be $\rho_{\Lambda} \sim R_{ge}^3$,
where $R_{ge}$ is Dirac's "big number" of the universe (the ratio of the
electromagnetic to gravitational forces between electrons and protons),
has the nice implication that the coincidence of our time with the time
when the universe gets a wish to accelerate is a necessity, basically for
the same reason why it is a necessity that the number of protons in the
present observable horizon is about the same as the square of $R_{ge}$,
hence saving us from anthropic headaches. So, perhaps this cosmological
constant is not so ugly after all.

\acknowledgements

  JM is supported in part by the Spanish National Program grants
AYA2006-06341 and AYA2006-15623-C02-01, and NASA grant NNG056H776.

\end{document}